\documentclass[letterpaper]{article} 
\usepackage[preprint]{aaai2027}  
\usepackage{amsmath}
\usepackage{amssymb}   
\usepackage{booktabs}  
\usepackage{graphicx}  
\usepackage{multirow}  
\usepackage{subcaption}
\usepackage[hyphens]{url}  
\usepackage{natbib}  
\usepackage{adjustbox}
\usepackage{caption} 

\newcommand{\resulttakeaway}[2]{%
    \par\smallskip
    \noindent
    \begingroup
    \setlength{\fboxsep}{5pt}%
    \colorbox{black!7}{%
        \parbox{\dimexpr\linewidth-2\fboxsep\relax}{%
            \small
            \textbf{Takeaway #1.}\enspace #2
        }%
    }%
    \endgroup
    \par\smallskip
}
\usepackage{fancyhdr}

\fancypagestyle{preprintfirstpage}{
  \fancyhf{}

  \fancyfoot[C]{\small\itshape Preprint version.}
}

\newcommand{\method}{\textsc{BeyondUncertainty}}
\title{Beyond Self-Knowledge: Propagating Uncertainty Across Reasoning and Retrieval in LLMs}

\author{
Chandan Kumar Sah,
Li Zhang,
Xiaoli Lian
}
\affiliations{
Beihang University, Beijing, China\\
\texttt{sahchandan98@buaa.edu.cn}\\
}

\begin{document}
\maketitle
\thispagestyle{preprintfirstpage}

\begin{abstract}
Retrieval-augmented generation improves knowledge-intensive question answering,
but indiscriminate retrieval can introduce irrelevant evidence and unnecessary
computation. We investigate whether verbalized confidence from black-box
language models can serve as an actionable signal for retrieval routing. Our
method, \method{}, first elicits a structured provisional answer and confidence
estimate, then applies a model-specific threshold selected on held-out
validation data and frozen before test evaluation. Low-confidence questions
receive top-$5$ TF--IDF retrieval followed by a second answer call, whereas
high-confidence questions return the provisional answer directly. We evaluate 27,000 policy instances across six QA benchmarks,
three model families, and three retrieval policies. \method{} achieves $0.483$ mean token-level F1, compared with $0.467$
for always retrieval and $0.401$ for no retrieval, while reducing retrieved
passages by $20.4\%$ relative to always retrieval. When matched on the number of questions routed to retrieval within
each dataset--model cell, it outperforms a post-hoc random allocation
in 17 of 18 settings, with an average gain of 0.024 F1. Although poorly calibrated as an absolute probability,
probe uncertainty modestly predicts question-level retrieval
benefit (AUROC $=0.628$). However, the additional probe increases total token usage
by $28.2\%$, revealing a trade-off between more selective evidence acquisition
and end-to-end token efficiency.
\end{abstract}

\noindent
\small
Our source code and supporting materials are available at\\
\url{https://github.com/Rocky5502/BeyondUncertainty_V1}.
\normalsize

\section{Introduction}
\label{sec:introduction}

Large language models (LLMs) can answer many factual questions from their
parametric knowledge, yet they may produce fluent but unsupported responses
when relevant knowledge is missing, outdated, or difficult to compose.
Retrieval-augmented generation (RAG) mitigates this limitation by grounding
generation in external evidence \cite{lewis2020rag}. Retrieval is not uniformly
beneficial, however: invoking a retriever for every question increases context
length and latency, while irrelevant or conflicting passages may distract the
generator and degrade answer quality
\cite{moskvoretskii2025adaptive,su2024dragin,shi2023distracted,
yoran2024robust}. Adaptive RAG therefore seeks to determine when external evidence is likely to
be useful. Existing methods base this decision on query complexity,
token-level confidence, reflection signals, internal model states, or estimated
information need
\cite{jeong2024adaptiverag,asai2023selfrag,jiang2023flare,
su2024dragin}. In parallel, research on LLM uncertainty examines
verbalized confidence, sampling consistency, and semantic entropy as indicators
of answer reliability
\cite{kadavath2022mostly,xiong2023uncertainty,
kuhn2023semantic,farquhar2024semanticentropy}. These directions motivate a
practically important question: can an uncertainty signal available through a
black-box API be transferred from a structured pre-retrieval probe into a
downstream retrieval decision, and can it identify which questions benefit
from evidence rather than merely reducing retrieval frequency?

We investigate this question using a deliberately simple and model-agnostic
controller. For each question, the LLM produces a structured provisional
answer, a verbalized confidence estimate, and a concise state summary. Routing
is determined solely by confidence and requires no access to hidden
chain-of-thought, logits, internal activations, or model fine-tuning. A
model-specific threshold, selected on held-out validation data and frozen
before test evaluation, determines whether the system returns the provisional
answer directly or retrieves the top-$5$ passages and performs a separate
final-answer call. This design isolates the operational value of black-box
confidence while remaining applicable to heterogeneous proprietary model
families.

We address three research questions:
\begin{itemize}
    \item \textbf{RQ1:} Does probe uncertainty predict which questions benefit
from retrieval, and how does this relate to final-answer quality?
    \item \textbf{RQ2:} How does confidence-guided routing trade answer quality,
    retrieved evidence, and gateway-reported token usage?
    \item  \textbf{RQ3:} Does confidence-guided routing outperform random
allocation when both route the same number of questions to retrieval
within each dataset--model cell?
\end{itemize}

Our contributions are threefold. First, we introduce a model-agnostic black-box
routing interface that propagates an observable confidence signal into an
external evidence-acquisition decision. Second, we conduct a paired evaluation comprising 27,000 policy
instances across six single- and multi-hop QA benchmarks and three
model families, retaining API and structured-output failures rather
than filtering them post hoc. Third, through a route-count-matched control, we show that confidence
provides question-level allocation value beyond retrieval frequency
alone. Our findings also expose important
limitations: verbalized confidence is poorly calibrated in absolute terms, the
additional probe makes the method retrieval-saving but not token-saving, and
improvements over always retrieval vary across dataset--model settings.

\section{Related Work}
\label{sec:related}

\paragraph{Adaptive retrieval.}
RAG combines generation with external evidence \cite{lewis2020rag}. IRCoT
interleaves retrieval and reasoning \cite{trivedi2022ircot}; FLARE and DRAGIN
trigger retrieval from token uncertainty or information need
\cite{jiang2023flare,su2024dragin}; Adaptive-RAG predicts query complexity
\cite{jeong2024adaptiverag}; and Self-RAG and SeaKR use reflection or
self-aware knowledge signals \cite{asai2023selfrag,yao2025seakr}. AdaRAGUE
compares a broad family of adaptive strategies and shows that simple
uncertainty signals can be competitive \cite{moskvoretskii2025adaptive}. Our
focus is narrower and complementary. Long-tail knowledge motivates retrieving
only when parametric memory is unreliable \cite{mallen2023trust}; RECOMP
selectively suppresses unhelpful retrieved text \cite{xu2024recomp}; and
robustness studies characterize how irrelevant context can reduce answer
quality \cite{shi2023distracted,yoran2024robust}. Recent work compares uncertainty-based routing, estimates retrieval need for
black-box models, and uses confidence to allocate search-agent computation
\cite{moskvoretskii2025adaptive,lin2025scaar,ou2026browseconf}. We isolate a
single verbalized-confidence signal across heterogeneous APIs and evaluate its
routing value, passage savings, and end-to-end token cost.

\paragraph{Uncertainty and calibration.}
LLM self-knowledge has been studied through answer verification and elicited
probabilities \cite{kadavath2022mostly}. Black-box estimators include
verbalized confidence, sampling agreement, and semantic dispersion
\cite{xiong2023uncertainty,lin2024generating,kuhn2023semantic,
farquhar2024semanticentropy,manakul2023selfcheckgpt}. Prior work also finds that
generative QA probabilities and verbalized certainty can be miscalibrated
\cite{jiang2021know,mielke2022linguistic,tian2023calibration,
yuan2024blackbox}. Because confidence can rank predictions while
remaining probabilistically miscalibrated, we separate \emph{ordinal utility}
from \emph{absolute calibration} and report both rank association and expected
calibration error (ECE) \cite{guo2017calibration}.
We provide a controlled cross-provider study demonstrating the operational
value of black-box confidence for retrieval allocation across answer quality,
passage use, token cost, failures, and matched routing decisions.
\section{Black-Box Confidence Routing}
\label{sec:method}

\begin{figure*}[t]
    \centering
    \includegraphics[width=0.88\textwidth]{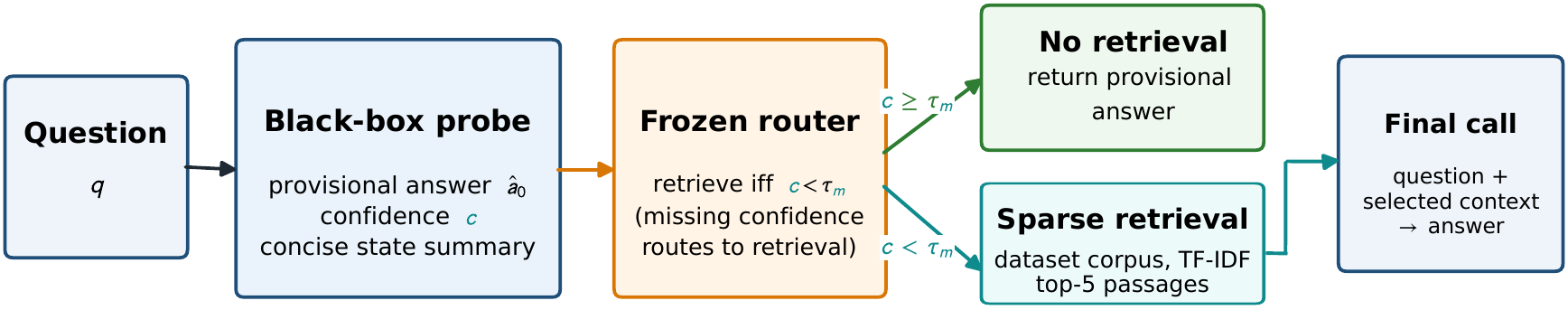}
    \caption{Implemented \method{} pipeline. A structured black-box probe
    produces a provisional answer and confidence. A validation-frozen,
    model-specific threshold routes the question either directly to that answer
    or through top-$5$ sparse retrieval and a second answer call.}
    \label{fig:framework}
\end{figure*}

Given question $q$, a black-box model first returns the observable state
\begin{equation}
s_0=(\hat{a}_0,c_0,d_0,m_0),
\end{equation}
where $\hat{a}_0$ is a provisional answer, $c_0\in[0,1]$ is verbalized
confidence, $d_0$ is the model's suggested action, and $m_0$ is a concise state
summary. The prompt requests compact JSON and explicitly asks for a summary
rather than hidden chain-of-thought. We define uncertainty as $u_0=1-c_0$.

For model family $m$, the frozen router is
\begin{equation}
R(q,m)=\mathbb{I}[c_0<\tau_m],
\label{eq:router}
\end{equation}
with missing or unparseable confidence conservatively mapped to retrieval. If
$R=0$, the system returns $\hat{a}_0$. If $R=1$, a deterministic sparse
retriever uses the original question as the query, returns up to five passages,
and the same model receives $q$ together with those passages in a final call:
\begin{equation}
P=\operatorname{TopK}(q,\mathcal{D}_j,5),\qquad
\hat{a}=\operatorname{LLM}(q,P).
\end{equation}
Here $\mathcal{D}_j$ is the fixed corpus for dataset $j$. The model's
suggested action $d_0$ is logged but does not determine routing, which
keeps the decision rule reproducible and isolates the role of confidence.
Figure~\ref{fig:framework} summarizes the implemented workflow. The threshold has a decision-theoretic interpretation. Let $Y_0$ and $Y_1$
denote answer utility without and with retrieval, and let $C_R$ denote
retrieval cost. A cost-sensitive controller retrieves when
\begin{equation}
\mathbb{E}[Y_1-Y_0\mid c_0,q] > \lambda C_R,
\label{eq:utility}
\end{equation}
where $\lambda$ encodes the value assigned to resource use. If expected
retrieval benefit decreases with confidence, a threshold in $c_0$ provides
a reduced-form approximation to this rule. We neither assume nor prove
perfect monotonicity. Instead, RQ1 tests whether confidence orders observed
quality, RQ2 reports quality and cost separately rather than fixing
$\lambda$, and RQ3 evaluates whether the induced allocation outperforms a
route-count-matched random allocation. This is a single-decision form of uncertainty propagation: an uncertainty
estimate produced during provisional reasoning is consumed by a downstream
retrieval component. It does \emph{not} implement iterative uncertainty
trajectories, learned routing, multi-step early stopping, or selective
abstention; these remain directions for future work.

\section{Experimental Design}
\label{sec:setup}

\paragraph{Datasets and splits.}
We evaluate six question-answering benchmarks: Natural Questions (NQ)
\cite{kwiatkowski2019natural}, HotpotQA \cite{yang2018hotpotqa},
2WikiMultiHopQA \cite{ho2020twowiki}, SQuAD \cite{rajpurkar2016squad},
TriviaQA \cite{joshi2017triviaqa}, and MuSiQue
\cite{trivedi2022musique}. Fixed question identifiers select 500 test examples
from each dataset, yielding 3,000 unique questions. Prompts and model-specific
routing thresholds are selected using held-out development data and frozen
before test evaluation. Test identifiers are not used for prompt design or
threshold selection.

\paragraph{Models and decoding.}
We access one gateway-reported alias from each of three model families through
the same OpenAI-compatible endpoint between July 8 and 11, 2026.
Table~\ref{tab:models} reports the aliases and frozen thresholds. A seed of
1337 is requested where supported, although bitwise determinism is not assumed
across providers. Gemini and Claude use temperature $0$ and a maximum output
length of 2,000 tokens. The OpenAI endpoint rejects temperature $0$; we
therefore use its gateway default and a 4,000-token output limit. Because none
of the evaluated endpoints exposes usable token log probabilities, verbalized
confidence provides the common black-box uncertainty signal.

\begin{table}[t]
\centering
\small
\setlength{\tabcolsep}{3.5pt}
\begin{tabular}{lll}
\toprule
Model family & Gateway-reported alias & $\tau_m$ \\
\midrule
OpenAI & \texttt{gpt-5.5} & 1.0 \\
Gemini & \texttt{gemini-3.5-flash} & 1.0 \\
Claude & \texttt{claude-sonnet-5} & 0.9 \\
\bottomrule
\end{tabular}
\caption{Gateway-reported model aliases and validation-frozen routing
thresholds. Retrieval is triggered when $c_0<\tau_m$.}
\label{tab:models}
\end{table}

The aliases are recorded exactly as returned by the gateway and should not be
interpreted as independent verification of direct-provider model versions.

\paragraph{Retrieval corpus and ranking.}
Table~\ref{tab:corpus-size} reports the number of deduplicated
title--passage documents in each dataset-specific retrieval corpus.
The operational backend is a deterministic TF--IDF retriever using
English stop-word removal and word uni- and bigrams. For each
benchmark, we construct a dataset-specific corpus from deduplicated
title--passage pairs in the processed development and test contexts.
The evaluation therefore measures routing over a controlled local
corpus rather than retrieval from a production-scale Wikipedia index.
All policies use the same corpus, preprocessing procedure, ranking
function, and retrieval depth.

\begin{table}[t]
\centering
\small
\setlength{\tabcolsep}{4.2pt}
\begin{tabular}{lr@{\hspace{10pt}}lr}
\toprule
Dataset & Documents & Dataset & Documents \\
\midrule
NQ       & 17,687 & SQuAD    & 14,369 \\
HotpotQA &  9,823 & TriviaQA & 14,696 \\
2Wiki    &  6,299 & MuSiQue  & 11,475 \\
\bottomrule
\end{tabular}
\caption{Number of deduplicated title--passage documents in each controlled
dataset-specific retrieval corpus.}
\label{tab:corpus-size}
\end{table}

\paragraph{Structured probe and frozen decisions.}
The probe requests a compact JSON object containing a provisional answer,
verbalized confidence, a suggested retrieval action, and a concise state
summary of at most two sentences. The prompt prohibits Markdown fences and
requests only a task-relevant summary rather than hidden chain-of-thought.
The model's suggested action is recorded for analysis but does not affect
routing. Model-specific thresholds are selected according to validation F1 and
then frozen at $1.0$ for OpenAI and Gemini and $0.9$ for Claude. No prompt,
threshold, retrieval depth, or routing parameter is modified after test
evaluation begins. Requests are indexed by deterministic configuration hashes. Interrupted runs
therefore reuse completed requests rather than issuing the same paid call
again.

\paragraph{Policies and evaluation scale.}
We execute three primary policies for every question--model pair.
\textsc{No Retrieval} performs one answer call without external context.
\textsc{Always Retrieve} retrieves the top-$5$ passages before one answer
call. \method{} follows Equation~\ref{eq:router}: it performs one probe call,
returns the provisional answer when retrieval is not triggered, and otherwise
performs top-$5$ retrieval followed by a second answer call. The resulting grid contains
$3{,}000\times3\times3=27{,}000$ executed policy rows, corresponding to
9,000 paired question--model instances. We additionally construct a deterministic route-count-matched random
control and a non-deployable oracle from paired no-retrieval and
always-retrieval outputs. These post-hoc
controls require no additional model calls. The static policies use one generation call per example. Confidence routing
uses one probe call when it does not retrieve and two calls when retrieval is
triggered. Retrieved-passage counts therefore measure evidence demand, whereas
total token counts capture the controller's additional inference overhead.
We consequently distinguish retrieval savings from total generation cost.

\paragraph{Metrics and failure policy.}
The primary quality metric is normalized token-level F1; Exact Match
(EM) is reported in the supplement. Retrieval use is measured by the
number of returned passages, and token usage sums gateway-reported
input and output tokens across all calls, including the probe. API and
structured-output parsing failures remain in the evaluation denominator
and receive zero quality scores; any raw score from malformed text is
retained only for provenance. Within each of the 18 dataset--model cells, we conduct two-sided paired
Wilcoxon signed-rank tests for always retrieval versus no retrieval,
confidence routing versus no retrieval, and confidence routing versus
always retrieval. The resulting 54 tests are jointly corrected using
Holm's procedure \cite{holm1979simple}. Full per-cell results, adjusted
$p$-values, prompts, corpus details, and failure counts are provided in
the supplement.

\paragraph{Route-count-matched and oracle controls.}
Within each dataset--model cell, a stable SHA-256 ordering selects
exactly the same number of questions for retrieval as
BeyondUncertainty. Selected examples use their paired
always-retrieval outputs, whereas unselected examples use their
paired no-retrieval outputs. The oracle selects the better of the two
observed static-policy outcomes for each example. Both controls are
deterministic and post hoc and require no additional model calls. The route-count-matched control tests whether confidence selects more
useful questions than random routing at the same route count. Because
some retrieval calls return fewer than five passages, equal route
counts do not imply identical returned-passage counts. Moreover, the
control does not reproduce BeyondUncertainty's structured probe or
two-stage execution path; it therefore evaluates allocation rather
than a fully execution-matched causal effect.

\section{Results}
\label{sec:results}

\begin{figure*}[t]
    \centering
    \includegraphics[width=0.75\textwidth]
    {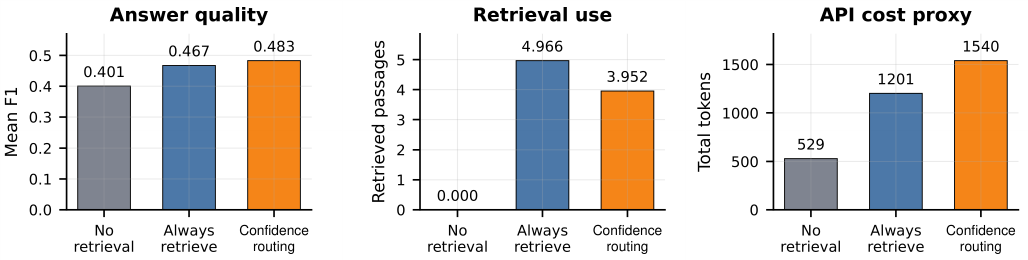}
    \caption{Aggregate answer quality and resource usage across
27,000 policy instances. Confidence routing achieves the highest
mean F1 and retrieves fewer passages than always retrieval, but its
additional probe increases total token usage.}
    \label{fig:overview}
\end{figure*}

\subsection{RQ1: Does Confidence Predict Retrieval Benefit?}

Table~\ref{tab:confidence_results} distinguishes answer-quality
prediction from retrieval-benefit prediction. Probe confidence is
positively associated with final F1 across all three model families
($\rho=0.276$--$0.309$), but is poorly calibrated as an absolute
probability ($\mathrm{ECE}_5=0.426$--$0.528$).

We define observed retrieval benefit as
\begin{equation}
\Delta_{\mathrm{ret}}
=
\mathrm{F1}_{A}-\mathrm{F1}_{N},
\end{equation}
where $A$ and $N$ denote always retrieval and no retrieval,
respectively. Across 8,837 instances with valid probe confidence,
uncertainty $u=1-c_0$ shows a positive but modest association with
retrieval benefit
($\rho=0.154$, 95\% CI $[0.131,0.177]$) and predicts
$\Delta_{\mathrm{ret}}>0$ with AUROC $0.628$ and AUPRC $0.447$,
against a positive-benefit prevalence of $0.354$. Per-model AUROC
ranges from $0.588$ to $0.633$.

\begin{figure}[t]
    \centering
    \includegraphics[width=0.98\columnwidth]
    {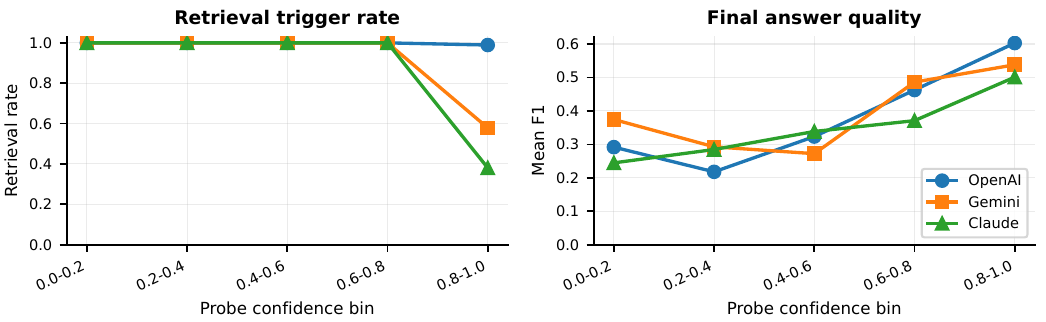}
    \caption{Retrieval rate and final F1 across probe-confidence bins.
    Confidence is poorly calibrated in absolute terms but generally orders
    questions by final answer quality.}
    \label{fig:confidence}
\end{figure}

The result is stable when failed static-policy pairs are excluded:
$\rho=0.150$, AUROC $=0.629$, and AUPRC $=0.448$. Mean confidence is
$0.849$, $0.875$, and $0.655$ for OpenAI, Gemini, and Claude,
respectively, compared with exact-match rates of $0.364$, $0.375$,
and $0.232$. Figure~\ref{fig:confidence} nevertheless shows that
higher-confidence bins generally attain higher final F1. Thus,
verbalized confidence is more useful as an ordinal routing signal than
as a calibrated estimate of correctness or retrieval benefit.

\begin{table}[t]
\centering
\small
\setlength{\tabcolsep}{3.5pt}
\renewcommand{\arraystretch}{1.08}
\begin{tabular}{lrrrr}
\toprule
\textbf{Model} &
$\boldsymbol{\rho}(c_0,\mathrm{F1})$ &
$\mathbf{ECE}_5$ &
$\boldsymbol{\rho}(u,\Delta_{\mathrm{ret}})$ &
\textbf{AUROC} \\
\midrule
OpenAI & 0.276 & 0.487 & 0.158 & 0.633 \\
Gemini & 0.300 & 0.528 & 0.126 & 0.606 \\
Claude & 0.309 & 0.426 & 0.168 & 0.588 \\
\bottomrule
\end{tabular}
\caption{Decision utility and calibration of probe confidence.
Here, $u=1-c_0$ and
$\Delta_{\mathrm{ret}}=\mathrm{F1}_{A}-\mathrm{F1}_{N}$.
AUROC measures prediction of strictly positive retrieval benefit.}
\label{tab:confidence_results}
\end{table}

\resulttakeaway{1}{Probe uncertainty modestly but consistently predicts question-level
retrieval benefit across model families, while verbalized confidence
remains poorly calibrated. It is therefore more useful as an ordinal
routing signal than as a probability of correctness or retrieval
benefit.}

\subsection{RQ2: What Are the Quality--Efficiency Trade-offs?}
\label{sec:rq2}

Table~\ref{tab:main-results} reports the aggregate policy comparison,
while Figure~\ref{fig:overview} summarizes answer quality,
retrieval use, and total token usage across the three executed policies.
\method{} achieves a mean F1 of $0.483$, improving over no retrieval by
$0.082$ absolute F1 and over always retrieval by $0.016$. It returns
$3.952$ passages per instance, compared with $4.966$ for always retrieval,
corresponding to a $20.4\%$ reduction. 
The improvement in retrieval allocation does not translate into lower total
token usage. The confidence probe increases mean token usage from $1201.5$ for
always retrieval to $1539.8$, an increase of $28.2\%$. Consequently,
\method{} is \emph{retrieval-saving} but not \emph{token-saving} under the
evaluated implementation. Table~\ref{tab:cell-results} shows that the aggregate improvement is
heterogeneous. Confidence routing numerically outperforms always retrieval in
16 of the 18 dataset--model cells, with the two exceptions occurring for
Gemini on NQ and SQuAD. However, after a single Holm correction across all 54
paired tests, only three confidence-versus-always comparisons remain
significant: NQ--OpenAI and SQuAD--Claude favor confidence routing, whereas
NQ--Gemini favors always retrieval.

\begin{table}[t]
\centering
\small
\setlength{\tabcolsep}{2.80pt}
\begin{tabular}{lrrr}
\toprule
Policy & F1 $\uparrow$ & Tokens / inst. $\downarrow$ &
Passages / inst. $\downarrow$ \\
\midrule
No retrieval          & 0.401 &  529.1 & 0.000 \\
Always retrieve       & 0.467 & 1201.5 & 4.966 \\
Random matched routes & 0.459 & 1036.5 & 3.927 \\
\textbf{\method{}}    & \textbf{0.483} & 1539.8 & 3.952 \\
Oracle no/always route & 0.523 & 1054.5 & 3.968 \\
\bottomrule
\end{tabular}
\caption{Aggregate comparison over 9,000 paired question--model
instances. The random control matches the number of questions routed
to retrieval within each dataset--model cell. Random and oracle
results are deterministic post-hoc recombinations of static-policy
outputs; their token counts exclude an independently executed routing
probe.}
\label{tab:main-results}
\end{table}

\begin{table*}[t]
\centering
\scriptsize
\setlength{\tabcolsep}{4.4pt}
\begin{tabular}{llrrrrr@{\hspace{12pt}}llrrrrr}
\toprule
Dataset & Family & F1-N & F1-A & F1-S &
$\Delta$(S--A) & Route &
Dataset & Family & F1-N & F1-A & F1-S &
$\Delta$(S--A) & Route \\
\midrule
NQ & OpenAI & 0.261 & 0.296 & \textbf{0.330} & +0.034$^*$ & 0.988 &
SQuAD & OpenAI & 0.288 & 0.446 & \textbf{0.463} & +0.017 & 0.998 \\

NQ & Gemini & 0.319 & \textbf{0.415} & 0.368 & -0.046$^*$ & 0.454 &
SQuAD & Gemini & 0.219 & \textbf{0.458} & 0.441 & -0.016 & 0.842 \\

NQ & Claude & 0.193 & 0.252 & \textbf{0.256} & +0.003 & 0.620 &
SQuAD & Claude & 0.171 & 0.333 & \textbf{0.356} & +0.024$^*$ & 0.924 \\
\midrule

Hotpot & OpenAI & 0.575 & 0.636 & \textbf{0.648} & +0.013 & 0.998 &
Trivia & OpenAI & 0.744 & 0.746 & \textbf{0.763} & +0.016 & 0.958 \\

Hotpot & Gemini & 0.549 & 0.589 & \textbf{0.600} & +0.011 & 0.584 &
Trivia & Gemini & 0.756 & 0.742 & \textbf{0.765} & +0.023 & 0.190 \\

Hotpot & Claude & 0.320 & 0.443 & \textbf{0.459} & +0.016 & 0.824 &
Trivia & Claude & 0.606 & 0.599 & \textbf{0.635} & +0.036 & 0.322 \\
\midrule

2Wiki & OpenAI & 0.622 & 0.651 & \textbf{0.666} & +0.015 & 1.000 &
MuSiQue & OpenAI & 0.351 & 0.416 & \textbf{0.417} & +0.001 & 0.968 \\

2Wiki & Gemini & 0.530 & 0.505 & \textbf{0.552} & +0.047 & 0.848 &
MuSiQue & Gemini & 0.256 & 0.292 & \textbf{0.330} & +0.039 & 0.848 \\

2Wiki & Claude & 0.320 & 0.377 & \textbf{0.415} & +0.038 & 0.948 &
MuSiQue & Claude & 0.129 & 0.212 & \textbf{0.228} & +0.016 & 0.918 \\
\bottomrule
\end{tabular}
\caption{Per-cell primary-policy results. N, A, and S denote no retrieval,
always retrieval, and confidence routing, respectively. Route is the fraction
of questions assigned to retrieval by S. A star denotes a significant S--A
difference after one joint Holm correction across all 54 paired tests. Bold
indicates the highest observed F1 in each dataset--model cell.}
\label{tab:cell-results}
\end{table*}

Three patterns are particularly important. First, retrieval is not uniformly
beneficial: always retrieval performs worse than no retrieval for Gemini on
2Wiki and TriviaQA and for Claude on TriviaQA. Second, the greatest
selectivity appears where confidence produces a broad routing distribution.
Gemini retrieves for only $19.0\%$ of TriviaQA and $45.4\%$ of NQ questions,
while Claude retrieves for $32.2\%$ of TriviaQA questions. Third, OpenAI routes
at least $95.8\%$ of questions to retrieval in every dataset. Its behavior is
therefore close to always retrieval, leaving little opportunity for meaningful
retrieval allocation. The multiplicity-corrected tests support a deliberately narrow inferential
claim. Always retrieval and confidence routing each significantly outperform
no retrieval in 11 of 18 cells. In contrast, confidence routing is difficult
to distinguish from always retrieval after correction: two cells significantly
favor confidence routing, one favors always retrieval, and the remaining 15
show no significant difference. The aggregate gain should therefore be viewed
as an average improvement across heterogeneous settings rather than evidence
of universal dominance.

\resulttakeaway{2}{Confidence routing achieves the highest aggregate F1 and
reduces retrieved passages by $20.4\%$ relative to always retrieval. However,
its additional probe increases total token usage by $28.2\%$, and its
cell-level advantage over always retrieval is not statistically uniform.}

\subsection{RQ3: Does Confidence Improve Allocation at a Matched
Route Count?}
\label{sec:rq3}

\begin{figure}[t]
    \centering
    \includegraphics[width=0.84\columnwidth]
    {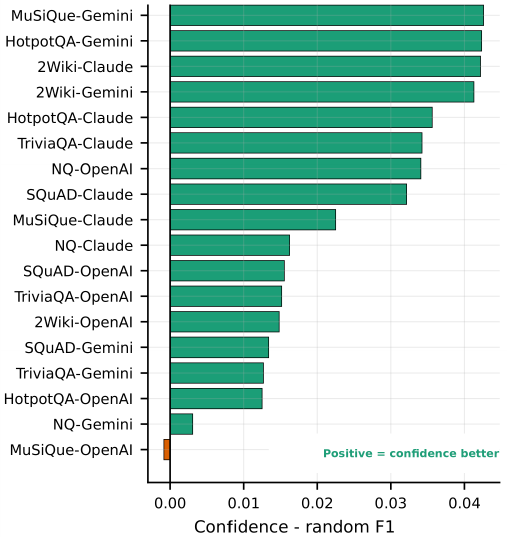}
   \caption{Per-cell F1 difference between confidence routing and random
routing when both route the same number of questions to retrieval
within each dataset--model cell. Positive values indicate that
confidence assigns retrieval to more useful questions.}
    \label{fig:matched-budget}
\end{figure}

The route-count-matched control retrieves for exactly the same
number of questions as \method{} within each
dataset--model cell, but chooses those questions using a stable
hash rather than confidence. This comparison separates the value
of which questions are retrieved from the effect of route count
alone. Confidence routing achieves 0.483 mean F1, compared with
0.459 for route-count-matched random routing, an average advantage
of 0.024 F1. It
outperforms the random allocation in 17 of the 18 cells shown in
Figure~\ref{fig:matched-budget}. The only exception is MuSiQue--OpenAI, for
which the difference is approximately $-0.001$.

The largest gains occur for MuSiQue--Gemini ($+0.043$), 2Wiki--Claude
($+0.042$), HotpotQA--Gemini ($+0.042$), and 2Wiki--Gemini ($+0.041$).
These settings combine nontrivial routing rates with substantial variation
in question-level retrieval benefit. Conversely, OpenAI's near-saturated
routing rate leaves little difference between confidence-based and random
allocations. The retrospective oracle chooses the better observed outcome between no
retrieval and always retrieval for every paired example. It reaches $0.523$
mean F1, leaving a $0.040$ gap above \method{}. This gap indicates meaningful
headroom for improved retrieval-benefit prediction. The route-count-matched comparison nevertheless has an important
limitation. It recombines outputs from the static policies and does
not reproduce the confidence controller's structured probe, prompt
path, or two-call execution. It therefore shows that the result is not
explained by route count alone, but it does not establish a fully
controlled causal effect of confidence. A directly executed randomized
router with the same probe and answer-call structure would provide a
stronger ablation.

\resulttakeaway{3}{At the same per-cell route count, confidence
routing outperforms random allocation in 17 of 18 dataset--model
cells and by 0.024 mean F1. Confidence therefore carries
question-level allocation value, although the post-hoc control does
not fully match the method's execution path.}

\section{Sensitivity, Robustness, and Interpretation}
\label{sec:robustness}

\begin{figure*}[t]
    \centering
    \begin{minipage}[c]{0.27\textwidth}
        \centering
        \includegraphics[width=\linewidth]
        {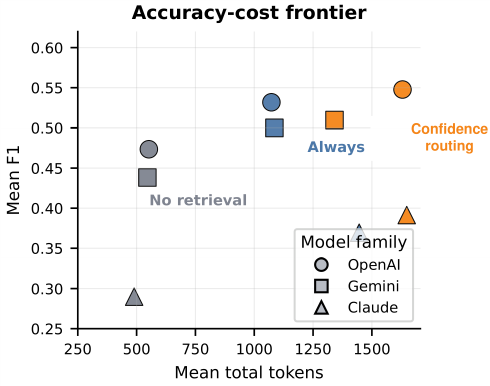}
    \end{minipage}
    \hfill
    \begin{minipage}[c]{0.68\textwidth}
        \centering
        \includegraphics[width=\linewidth]
        {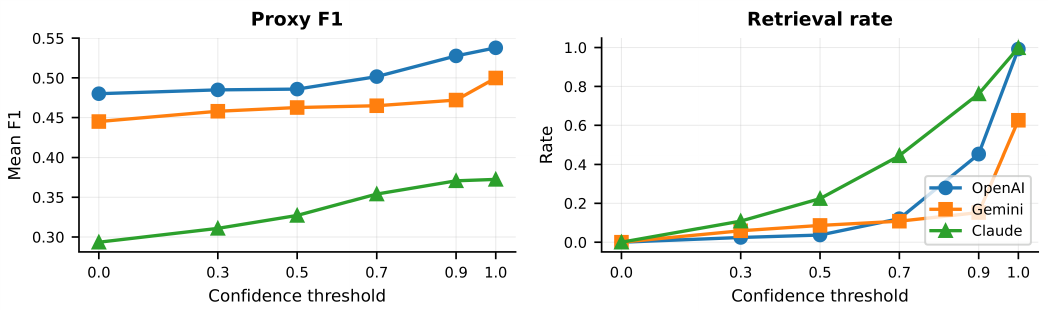}
    \end{minipage}
    \caption{Sensitivity and efficiency analyses.
    \textbf{Left:} Mean F1 versus gateway-reported token usage for each model
    family and retrieval policy.
    \textbf{Right:} Offline sensitivity to the routing threshold, constructed
    by recombining paired no-retrieval and always-retrieval outputs. Increasing
    the threshold generally routes more questions to retrieval and improves
    proxy F1, but also increases passage and token usage.}
    \label{fig:sensitivity}
\end{figure*}

\paragraph{Sensitivity to the routing threshold.}
We evaluate thresholds
$\tau\in\{0,0.3,0.5,0.7,0.9,1.0\}$ through an offline paired-output sweep.
For every probe with parseable confidence, the proxy uses the corresponding
always-retrieval output when $c_0<\tau$ and the no-retrieval output otherwise.
Observed probe tokens are then added to the selected branch. This procedure
preserves the paired static-policy outcomes but does not issue new model calls;
it should therefore be interpreted as a diagnostic sensitivity analysis rather
than an independently executed policy evaluation. Among usable probe outputs, setting $\tau=1.0$ produces proxy
F1/retrieval-rate pairs of $0.538/99.2\%$ for OpenAI, $0.500/62.6\%$ for
Gemini, and $0.373/99.9\%$ for Claude. The substantially different response
curves demonstrate that a common numerical threshold does not induce a common
retrieval budget across model families.
In the executed primary-policy runs, the validation-frozen thresholds produce
mean retrieval rates of $98.5\%$ for OpenAI, $62.8\%$ for Gemini, and
$75.9\%$ for Claude. Thus, the shared threshold of $1.0$ yields almost no
filtering for OpenAI but substantial filtering for Gemini. Routing behavior is
determined jointly by the threshold and the model-specific confidence
distribution; the numerical value of $\tau_m$ is not meaningful in isolation.

\paragraph{When does adaptivity help?}
The value of routing depends on whether retrieval has heterogeneous
question-level effects. Always retrieval performs worse than no retrieval for
Gemini on 2Wiki and TriviaQA and for Claude on TriviaQA, indicating that
retrieved context can sometimes interfere with an otherwise correct parametric
answer. Confidence routing outperforms both static policies in all three
settings. By contrast, OpenAI routes at least $95.8\%$ of questions to retrieval in every
dataset. Its realized behavior is therefore close to always retrieval, leaving
little opportunity for passage savings or meaningful question-level
selection. The aggregate benefit of adaptivity consequently depends on both
the usefulness of external evidence and the degree of variation in each
model's confidence distribution.

\begin{figure}[t]
    \centering
    \includegraphics[width=0.78\columnwidth]
    {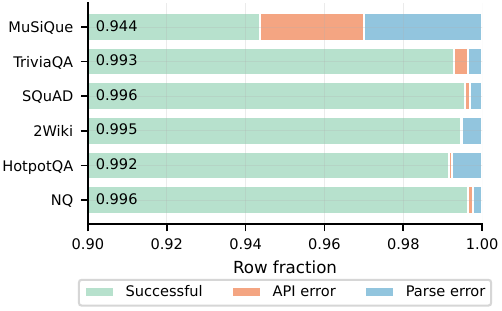}
    \caption{Coverage and retained failures across datasets. Successful,
    API-error, and structured-output parse-error rows partition the complete
    planned evaluation grid; failures remain in the scoring denominator.}
    \label{fig:coverage}
\end{figure}

\paragraph{Failure robustness and artifact completeness.}
The complete evaluation artifact contains all 27,000 planned
primary-policy rows, forming 9,000 complete no-retrieval,
always-retrieval, and confidence-routing triples.
Figure~\ref{fig:coverage} summarizes dataset-level coverage
and the retained API and structured-output parsing failures.
The artifact includes 152 API-error rows and 232 structured-output
parsing failures, all of which remain in the denominator and receive
zero answer-quality scores. Coverage exceeds $99\%$ for NQ, HotpotQA, 2Wiki, SQuAD, and TriviaQA.
MuSiQue has lower coverage at $94.4\%$, accounting for 119 API failures and
135 parsing failures. To test whether failure retention determines the main
result, we additionally compute scores over successful rows only. Mean F1
changes from $0.401$, $0.467$, and $0.483$ to $0.409$, $0.472$, and $0.488$
for no retrieval, always retrieval, and confidence routing, respectively.
The policy ordering and the rounded $0.016$ confidence-versus-always gap remain
unchanged. Five continuation records missing after an interrupted resume were recovered
from the preceding complete artifact. Their original failure outcomes were
preserved, and no answer or score was imputed. Row-level provenance, request
identifiers, parsing status, and complete error counts are included in the
released artifact and supplementary material.

\paragraph{Multiplicity-aware statistical evidence.}
Table~\ref{tab:test-summary} summarizes the paired tests after one joint Holm
correction across all 54 dataset--model comparisons. Always retrieval and
confidence routing each significantly outperform no retrieval in 11 of 18
cells. Confidence routing is more difficult to distinguish from always
retrieval: only three comparisons survive correction, with two favoring
confidence routing and one favoring always retrieval. This evidence supports a narrow interpretation. Confidence routing yields a
positive aggregate improvement, but the effect is distributed
heterogeneously across datasets and model families rather than appearing as a
uniform advantage in every cell.

\begin{table}[t]
\centering
\small
\setlength{\tabcolsep}{3.0pt}
\begin{tabular}{lrrrr}
\toprule
Comparison &
$\Delta$F1 &
Holm sig. &
Favor first &
Favor second \\
\midrule
A vs.\ N & +0.067 & 11/18 & 11 & 0 \\
S vs.\ N & +0.082 & 11/18 & 11 & 0 \\
S vs.\ A & +0.016 &  3/18 &  2 & 1 \\
\bottomrule
\end{tabular}
\caption{Summary of 54 two-sided paired Wilcoxon tests after
one joint Holm correction. N, A, and S denote no retrieval,
always retrieval, and confidence routing, respectively. The
aggregate difference is the mean F1 of the first policy minus
that of the second.}
\label{tab:test-summary}
\end{table}

\paragraph{Efficiency interpretation.}
Figure~\ref{fig:sensitivity} places the evaluated policies in the
answer-quality--token-usage plane. Confidence routing occupies the
highest-quality but highest-token operating point among the three executed
policies: it improves mean F1 from $0.467$ to $0.483$ relative to always
retrieval, while increasing gateway-reported token usage from $1201.5$ to
$1539.8$ per instance. At the same time, it reduces the average number of returned passages from
$4.966$ to $3.952$. These results reinforce the distinction between retrieval
efficiency and end-to-end inference efficiency. The current controller assigns
external evidence more selectively, but the additional black-box probe more
than offsets those passage savings in total token usage. The post-hoc oracle
should not be interpreted as a deployable point on this trade-off because it
selects between outcomes after observing both static-policy results.

\resulttakeaway{4}{Routing behavior is strongly model-dependent and cannot be
inferred from the numerical threshold alone. The principal findings remain
unchanged when failed requests are excluded: confidence routing improves the
allocation of retrieved evidence, but its additional probe prevents an
end-to-end token-efficiency gain.}

\section{Discussion and Limitations}
\label{sec:discussion}

Our results distinguish the decision utility of confidence from its absolute
calibration. Although verbalized confidence is overconfident, it still ranks
answer quality and improves retrieval allocation relative to a
route-count-matched random policy. This benefit is model-dependent and weakens
when confidence values cluster near the routing threshold.

Confidence routing reduces retrieved passages by $20.4\%$, but the additional
probe increases total token usage by $28.2\%$. The method therefore improves
evidence allocation rather than end-to-end token efficiency. Its main
limitations are the controlled TF--IDF corpus, unmatched execution paths,
time-specific gateway aliases, and heterogeneous cell-level significance.
Future work should explore cheaper probes, execution-path-matched controls,
and frozen open-domain retrieval indexes. Additional threats to validity are
provided in the supplementary material. \noindent\textit{Overall, black-box confidence improves retrieval allocation
under controlled QA conditions, but not end-to-end token efficiency.}

\paragraph{Ethical considerations.}
This study uses public QA benchmarks and involves no human participants or
newly collected personal data. Because verbalized confidence is poorly
calibrated and third-party gateways may raise privacy and reproducibility
concerns, our claims are limited to retrieval allocation and should not be
extended to safety-critical deployment without task-specific validation,
monitoring, and human oversight.
\section{Conclusion}
\label{sec:conclusion}

We presented \method{}, a black-box confidence-guided retrieval strategy for
question answering. Across 27,000 policy instances, BeyondUncertainty achieves the
highest aggregate F1 while reducing retrieved passages by 20.4\%
relative to always retrieval. Its advantage over route-count-matched random routing further indicates
that verbalized confidence provides useful question-level allocation signals,
despite poor absolute calibration. However, the additional probe increases
total token usage, so the current method improves retrieval allocation rather
than end-to-end inference efficiency. These findings motivate future work on
cheaper confidence estimators, execution-path-matched controls, and
open-domain retrieval settings.
\bibliography{references}

\end{document}